# Scaling Laws and Paradoxical Metastable States in Nanofilament Entropic Separation

Jose M. G. Vilar[1,2,*], J. Miguel Rubi[3], and Leonor Saiz[4,*]

[1] Biofisika Institute (CSIC, UPV/EHU), University of the Basque Country (UPV/EHU), P.O. Box 644, 48080 Bilbao, Spain

[2] IKERBASQUE, Basque Foundation for Science, 48011 Bilbao, Spain

[3] Departament de Fisica de la Materia Condensada, Universitat de Barcelona, Marti i Franques 1, 08028, Barcelona, Spain

[4] Department of Biomedical Engineering, University of California, 451 E. Health Sciences Drive, Davis, CA 95616, USA

[*] Correspondence to: j.vilar@ikerbasque.org or lsaiz@ucdavis.edu

## Abstract

Entropic forces play a fundamental role in nanoscale phenomena, from colloidal self-assembly to biomolecular disaggregation. Here, we develop an exact analytical theory and find general scaling laws for the entropic separation of tether-mediated nanofilament bundles, revealing that a single dimensionless parameter—the ratio of the excluded-volume radius to the tether length—dictates whether filaments are pushed apart or, contrary to the usual expectation, pulled together. This unexpected regime challenges the view that entropic forces invariably promote disaggregation, instead uncovering conditions under which the bundles can remain in attractive metastable states. Brownian dynamics simulations confirm this paradoxical effect, offering predictive insights for applications in biophysics, soft matter physics, and nanotechnology.

# I. Introduction

Entropic forces play a fundamental role in nature, influencing a wide range of phenomena at the nanoscale. They are central to understanding phenomena as diverse as the self-assembly of colloidal systems[1], the disassembly of biomolecular structures[2,3], the translocation of chains and polymers through narrow pores[4,5], and the transport of nanoparticles through non-uniform channels[6-8]. The key determinant is the interplay between confinement and the statistical properties of the system's components. Confinement restricts the available space for the system's degrees of freedom, leading to a reduction in the overall system entropy. In this context, we distinguish between confinement as the physical boundaries enclosing the system (e.g., rigid walls or tethers) and constraints as the specific geometric limitations these boundaries impose on the available phase space. The system then moves against the geometric constraints to reach a state of maximum entropy, which can manifest in multiple ways depending on the specific system and its degrees of freedom[9].

In the disaggregation of amyloid fibril bundles, consisting of misfolded protein aggregates implicated in neurodegenerative diseases like Alzheimer's and Parkinson's, the rate-limiting step is the separation of the filaments of the bundle before a sudden disassembly of the filaments[2,10,11]. The separation process is entirely driven by entropic forces through a recently uncovered mechanism, termed entropic separation, in which molecules, known as molecular chaperones[12], are tethered to each of the filaments[11]. Separation refers to the unmixing of two subsystems, each consisting of a filament with its tethered molecules, which ultimately repel each other. The tethered particles introduce steric hindrance, leading to a decrease in the overall entropy of the aggregated state. When the two subsystems are separated, the overall entropy increases because the steric hindrance decreases, resulting in the dissociation of the fibril bundle. The phenomenology, as characterized through Brownian dynamics, is that of a highly optimized system that exerts forces strong enough to overcome the otherwise stable lateral filament association[11].

There are substantial differences with other types of entropic processes. Unlike the prototypical confinement by rigid walls[9], here, the confinement is mediated by tethers, allowing the two subsystems to partially overlap. As the filament separation changes, the overlap can change continuously, so the two subsystems can transition from mixed (substantial overlap) to unmixed (little or no overlap) progressively, rather than via an abrupt partition. The mechanism also differs from that of depletion forces[13,14]. In such case, the entropy change originates within a fixed system volume; bringing objects together primarily changes the volume available to depletants by excluding them from the inter-object gap. In entropic separation, the tether-imposed domains and steric exclusion jointly define an effective accessible volume beyond the inter-object gap, and this accessible volume itself varies with filament separation, producing a more complex distance-dependent entropic interaction. Unlike the pulling forces of chain translocation inside cells[4], it does not rely on changes in the binding entropy of molecules to the chain. Furthermore, unlike polymer translocation[5], forces in entropic separation do not arise from changes in the available polymer conformations. Therefore, although there are common features, the inherent differences with other entropic processes make it difficult to fully grasp the potential of this new mechanism without a supporting theoretical background.

Here, we develop an exact analytical theory for the entropic separation of nanofilament bundles and compare the results with Brownian dynamics numerical experiments. In contrast to the highly optimized natural biological case, we uncover a range of behavior types, including paradoxical regimes in which the stability of the aggregated bundles is enhanced, promoting their persistence under non-equilibrium conditions rather than their disaggregation. Through general scaling laws, we found that the type of behavior is determined by just one dimensionless parameter.

Our analytical results provide an effective avenue to tackle the inherent complexity of tethering-mediated disaggregation processes. Understanding the conditions under which entropic forces promote or hinder disaggregation is crucial for developing targeted therapeutic strategies for neurodegenerative diseases. The ability of mechanisms from highly optimized biomolecular systems to exhibit novel behaviors beyond their natural operating conditions is also important in a broad range of fields, including biophysics[15], soft matter physics[16], targeted drug delivery[17], and nanoparticle assembly[18], all of which have developed methods to tether particles to modify and control the properties at the nanoscale.

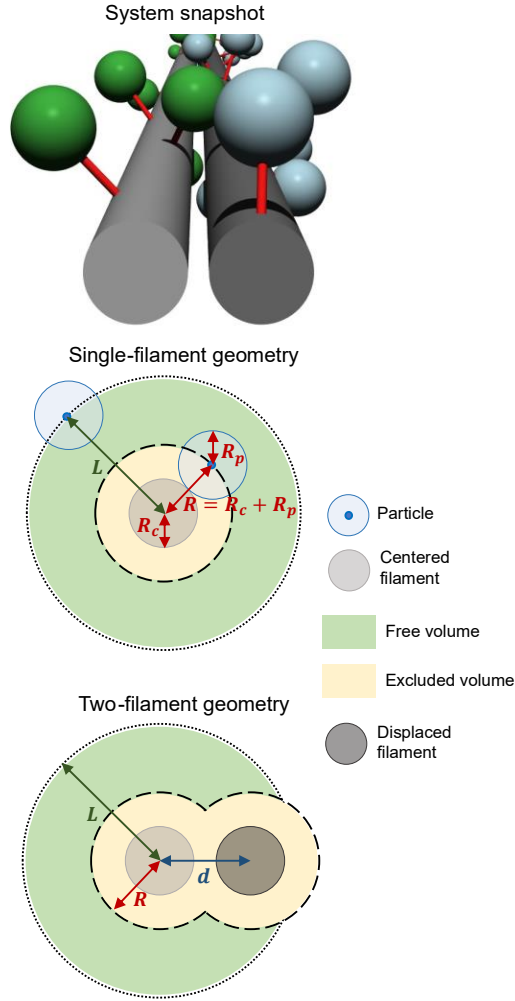

**Figure 1**. System description. The system comprises a nanofilament bundle described as two parallel cylinders with tethered spherical particles. (Top: System snapshot) Representative state of the Brownian dynamics system for two nanofilaments of radius $R_c = 2.5$ nm at a distance $d = 6$ nm between their axes, with particles of radius $R_p = 2.5$ nm attached to them through a tether of length $L = 10$ nm. (Middle: Single-filament geometry) Cross-section perpendicular to the filament axis showing the geometric constraints on the tethered particle. The particle center is restricted to lie within a distance $L$ from the tether origin and outside a distance $R = R_c + R_p$ from the axis of the filament (dashed circle), where $R$ is the excluded-volume radius. The accessible region for the particle center (green) is the annular area between these two boundaries. Two representative particle positions are shown: one tangent to the filament surface and one at the tether boundary. (Bottom: Two-filament geometry) Cross-section showing how a second filament, at a distance $d$ from the first, introduces an additional exclusion zone (yellow) that encroaches into the tether-accessible region, reducing the free volume (green) available to the particle. The entropic force arises from the change of the free volume with the inter-filament distance $d$.

## II. System characterization

We consider a nanofilament bundle described as two parallel cylinders with tethered spherical particles (Figure 1). The system parameters are the cylinder radius $R_c$, the distance $d$ between the axes of the cylinders, the particle radius $R_p$, and the effective tether length $L$ (defined as the maximum radial distance from the filament axis to the particle center). The motion of the center of the particle takes place between a distance $L$ from the tether origin and $R = R_c + R_p$, referred to as the excluded volume radius, along the axis of the filament. We consider the tether origins along the filament axis spaced a distance larger than $2(L + R_p)$ from each other so that tethered particles do not interact between them, but only with the filaments.

The cylinders representing the nanofilaments have a two-dimensional (2D) symmetry whereas the particles are restricted to moving within a three-dimensional (3D) sphere determined by the tether. To tackle these mixed geometric aspects, we consider first the motion of the particles in a 2D circle and subsequently, we integrate over the filament axis, from $-\sqrt{L^2 - R^2}$ to $\sqrt{L^2 - R^2}$ relative to the tether attachment point, to capture the overall 3D behavior. The 2D case is also representative because as $L$ gets closer to $R$, the 3D system behavior gets increasingly similar to the 2D case.

The entropy of the particle in 2D can be expressed as $S = k_B \ln\left(\frac{A_F}{\lambda^2}\right)$, where $k_B$ is the Boltzmann constant, $\lambda$ is a characteristic length scale of the system, and $A_F$ is the area the particle is allowed to move in. In 3D, it is given by $S = k_B \ln\left(\frac{V_F}{\lambda^3}\right)$, where $V_F$ is the volume instead of the area. The quantities $A_F$ and $V_F$ can be viewed as the free area and volume, respectively, available to the tethered particle. The entropic force per particle between filaments is related to the entropy as $f = T\frac{\partial S}{\partial d}$, where $T$ is the absolute temperature.

## III. Analytical results in two dimensions

In the 2D case, the tether length constrains the motion within a radius $L$ and area $A_L = \pi L^2$. Not all of this area is accessible because the filament the particle is attached to, referred to as the centered filament, excludes an area $A_R = \pi R^2$ (Figure 1).

The presence of another filament, referred to as displaced filament, introduces an additional excluded area that can overlap with the area allowed by the tether (Figure 2). These areas can be computed in terms of the radius $r$ and the height of the segment $h$ through the expression for the area of the circular segment

$$A_{CS}(r,h) = r^2 \arccos\left(1 - \frac{h}{r}\right) - (r-h)\sqrt{r^2 - (r-h)^2}.$$

(1)

In the case of the excluded area of the two filaments, they intersect at a distance $d/2$ along the axis across the two centers with the origin at the attachment point. This intersection leads to the heights of the circular segments of $R - \frac{d}{2}$ for the centered filament and $R + \frac{d}{2}$ for the displaced filament. Therefore, the additional excluded area $A_{RR}$ outside that of the centered filament is given by $A_{RR} = A_{CS}\left(R, R + \frac{d}{2}\right) - A_{CS}\left(R, R - \frac{d}{2}\right)$.

The excluded area of displaced filament that falls outside the area allowed by the tether can be computed by considering that both surfaces intersect at a position $x = \frac{d + L^2 - R^2}{2d}$. The heights of the circular segments are given by $R + d - x$ for the displaced filament and $L - x$ for the tether area. Therefore, the excluded area $A_{LR}$ that falls outside the allowed area is $A_{LR} = A_{CS}\left(R, R + d - \frac{d^2 + L^2 - R^2}{2d}\right) - A_{CS}\left(L, L - \frac{d^2 + L^2 - R^2}{2d}\right)$.

The overall excluded area within the tether length is $A_R + A_{RR} - A_{LR}$, and consequently, the free area is $A_F = A_L - A_R - A_{RR} + A_{LR}$, which leads explicitly to

$$A_F(L, R, d) = \pi(L^2 - R^2) - A_{CS}\left(R, R + \frac{d}{2}\right) + A_{CS}\left(R, R - \frac{d}{2}\right) + A_{CS}\left(R, R + d - \frac{d^2 + L^2 - R^2}{2d}\right)$$
$$- A_{CS}\left(L, L - \frac{d^2 + L^2 - R^2}{2d}\right).$$

(2)

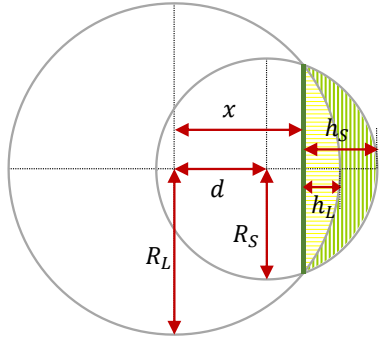

**Figure 2**. Geometric framework. The overlap of circles determines the available and excluded areas in 2D. The area of the small circle outside the large circle (green shade) is given by the area of the circular segment of radius $R_S$ and height $h_S$ (green and yellow shade) minus the area of the circular segment of radius $R_L$ and height $h_L$ (yellow shade). The two circles, at a distance $d$, intersect at $x = \frac{d^2 + R_L^2 - R_S^2}{2d}$, obtained from $R_L^2 - x^2 = R_S^2 - (x-d)^2$ since both segments have the same chord length, which leads to the dependence of $h_L = R_L - x$ and $h_S = d + R_L - x$ on the distance $d$ through $x$.

## IV. Analytical results in three dimensions

To analyze the 3D case, we implement a decomposition of the 3D volume as a continuum of planes perpendicular to the axis of the filaments. For a plane that is at a distance $l$ from the attachment point, the motion is restricted within a radius $\sqrt{L^2 - l^2}$ and an area $A_F(\sqrt{L^2 - l^2}, R, d)$. Therefore, the overall free volume is given by

$$V_F = \int_{-\sqrt{L^2 - R^2}}^{\sqrt{L^2 - R^2}} A_F\left(\sqrt{L^2 - l^2}, R, d\right) dl.$$

(3)

With this expression of $V_F$, we obtain the entropic force $f$ as a function of the distance $d$ and tether length $L$, as shown in the force landscape for a typical value of $R = 5\ nm$ (Figure 3). The force per particle exhibits an intricate behavior, which includes attractive, repulsive, and non-interacting domains. From the biological point of view, the optimal parameters are those that give a maximum repulsive force. The force landscape shows that this value is obtained for short tether lengths. Indeed, it has been observed that particles remain close to the filaments in the natural biological system[10]. In contrast to the requirements of the natural biological function, there are combinations of parameter values that lead to attractive forces. This situation would stabilize the bundle at short distances and potentially lower the escape rate from the bundled state[19].

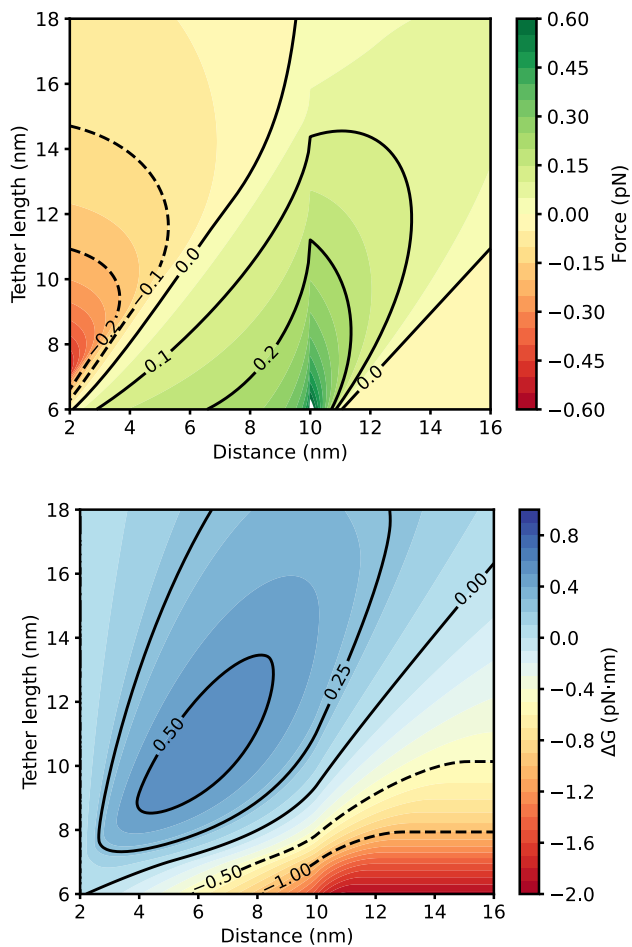

**Figure 3**. Force and energy landscapes with attractive and repulsive domains. The force per particle exhibits intricate behavior as a function of the distance and the tether length, as shown in the top panel for $R = 5\ nm$. This includes attractive, repulsive, and non-interacting domains, as delimited by the contour lines labeled 0.0. The corresponding free energy of separation per particle, as shown in the bottom panel for $R_c = 1\ nm$ and $R_p = 4\ nm$, remains positive in a large region of the phase space, becoming negative for small tether-length-to-distance ratios.

We also obtained the free energy of separation for the same conditions as those of the entropic force with $R_c = 1\ nm$ (Figure 3). The explicit value of $R_c$ is needed because, although the force does not depend on it, only on $R$, the separation free energy change does. Explicitly, the free energy change at a distance $d$ is given by $\Delta G(d) = TS(2R_c) - TS(d)$, which considers changes from the filaments closest possible distance $2R_c$ to a distance $d$. The corresponding free energy of separation per particle remains positive in a large region of the energy landscape, becoming negative only for small tether-length-to-distance ratios. This result indicates that the parameter values must lie in a narrow space for the system to achieve its intended biological function and that there is a high potential to influence the system's behavior.

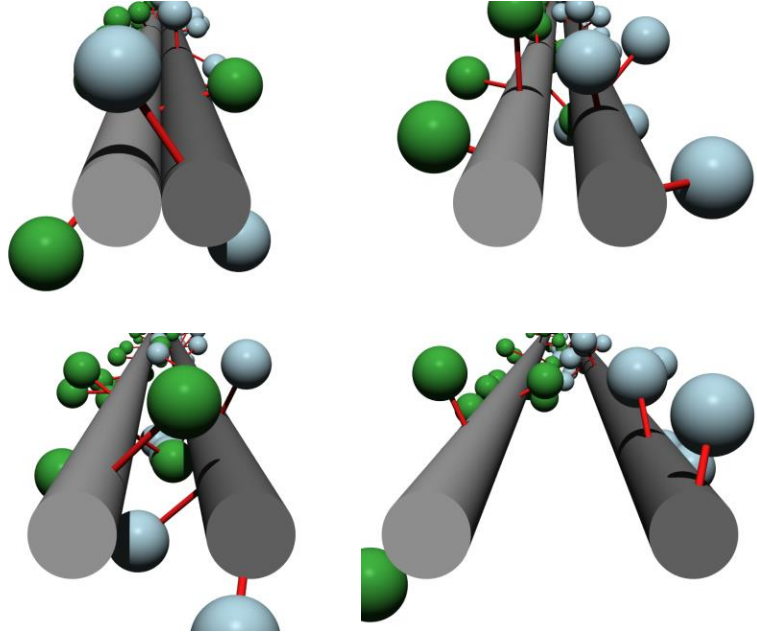

**Figure 4**. Representative Brownian-dynamics configurations of the coarse-grained two-filament system. Equilibrated snapshots are shown for filament axis–to–axis separations $d = 5$, 7, 10, and 15 nm (top left, top right, bottom left, bottom right, respectively). Filaments are modeled as fixed cylinders (gray, radius $R_c$). Tethered particles are hard spheres (radius $R_p$) colored by the filament to which they are attached (green vs light blue). The snapshots illustrate how increasing $d$ progressively relieves steric confinement and inter-filament coupling.

## V. Brownian dynamics

To validate the approach, we performed Brownian dynamics numerical experiments[20]. Filaments were modeled as two fixed, parallel cylinders of radius $R_c$ with axes along $\hat{\mathbf{z}}$ and separated by an axis-to-axis distance $d$. Each filament carries tethered spherical particles of radius $R_p$ attached at prescribed axial locations. Attachment points were chosen to be separated by more than $2(L + R_p)$ along $\hat{\mathbf{z}}$, so that particles tethered to different origins do not interact with one another; consequently, the force is additive and can be reported per particle, as in the theory.

The position $\mathbf{r}_i(t)$ of particle $i$ evolves according to the discretized overdamped Langevin (Brownian dynamics) equation

$$\Delta \mathbf{r}_i = \frac{D_i}{k_B T} \mathbf{F}_i \Delta t + \sqrt{2 D_i \Delta t}\, \mathbf{G}_i,$$

(4)

where $\Delta \mathbf{r}_i = \mathbf{r}_i(t + \Delta t) - \mathbf{r}_i(t)$, $D_i$ is the translational diffusion coefficient, $\mathbf{F}_i$ is the total deterministic force on the particle (arising solely from constraint forces described below), $k_B$ is Boltzmann's constant, $T$ is temperature, and $\mathbf{G}_i$ is a vector of independent, normally distributed random variables with zero mean and unit variance. We used the Stokes–Einstein relation $D_i = k_B T/(6\pi\eta R_p)$. For water at 25°C[21], this gives $D = 0.098$ nm$^2$ns$^{-1}$ for $R_p = 2.5$ nm, i.e. $D \approx 0.245/R_p$ in nm$^2$ns$^{-1}$ with $R_p$ in nm.

The force $\mathbf{F}_i$ acting on a particle comprises hard-core and tethering interactions[11]. Hard-core interactions between a particle $i$ and a filament $j$ are perpendicular to the filament axis and are described by

$$\mathbf{F}_{ij}^{(\text{fil})} = H\big(R_i + R_j - \| \mathbf{P}(\mathbf{r}_i - \mathbf{w}_j) \|\big) \frac{\mathbf{P}(\mathbf{r}_i - \mathbf{w}_j)}{\| \mathbf{P}(\mathbf{r}_i - \mathbf{w}_j) \|},$$

(5)

where $H(r)$ is zero for $r < 0$ and infinity otherwise, $\mathbf{P}(\mathbf{r}) = \mathbf{r} - (\mathbf{r} \cdot \hat{\mathbf{z}})\hat{\mathbf{z}}$ is the projection of the vector $\mathbf{r}$ onto the plane perpendicular to the filament axis direction $\hat{\mathbf{z}}$, and $\mathbf{w}_j$ is any position along the axis of the filament $j$. Tethering restricts how far a particle can be from its attachment point and is described through

$$\mathbf{F}_{ij}^{(\text{teth})} = -H\big(\| \mathbf{r}_i - \mathbf{r}_j \| - L\big) \frac{\mathbf{r}_i - \mathbf{r}_j}{\| \mathbf{r}_i - \mathbf{r}_j \|},$$

(6)

where $L$ is the tether length and $\mathbf{r}_j$ is the position of the attachment point on the filament. The total deterministic force is $\mathbf{F}_i = \sum_j \left( \mathbf{F}_{ij}^{(\text{fil})} + \mathbf{F}_{ij}^{(\text{teth})} \right)$. Since both the coarse-grained spherical particles and the tether have radial symmetry, only translational degrees of freedom are considered.

To implement the simulations numerically, the function $H(r)$ is approximated as $H(r) = F_W\,\Theta(r)$, where $\Theta$ is the Heaviside unit step function and $F_W$ is the force intensity. Within this framework, the average excess distance into the hard-core region or outside the tether length is $\Delta r \simeq k_B T/F_W$. The value of $F_W$ should be chosen large enough so that $\Delta r \ll R_p$ and $\Delta r \ll L$. We used $F_W = 100$ pN, which leads to $\Delta r \simeq 0.04$ nm, keeping the offset length below 1% of the center-to-center distances in a collision. The time step $\Delta t$ must be chosen small enough to be consistent with the continuous limit, satisfying $\Delta t \ll \Delta r^2/2D$, which leads to $\Delta t \ll 0.008$ ns for $F_W = 100$ pN. In practice, we used $\Delta t = 0.001$ ns.

The tethered particles were initialized with random orientations around the external side of their respective filaments, perpendicular to the filament axis. An equilibration phase of $2\,\mu s$ was performed in which the force intensity was increased linearly from $F_W = 1$ pN to $F_W = 100$ pN and, concomitantly, the time step was decreased quadratically from $\Delta t = 0.01$ ns to $\Delta t = 0.001$ ns. This gradual protocol ensures that any steric clashes present in the initial conditions are resolved smoothly.

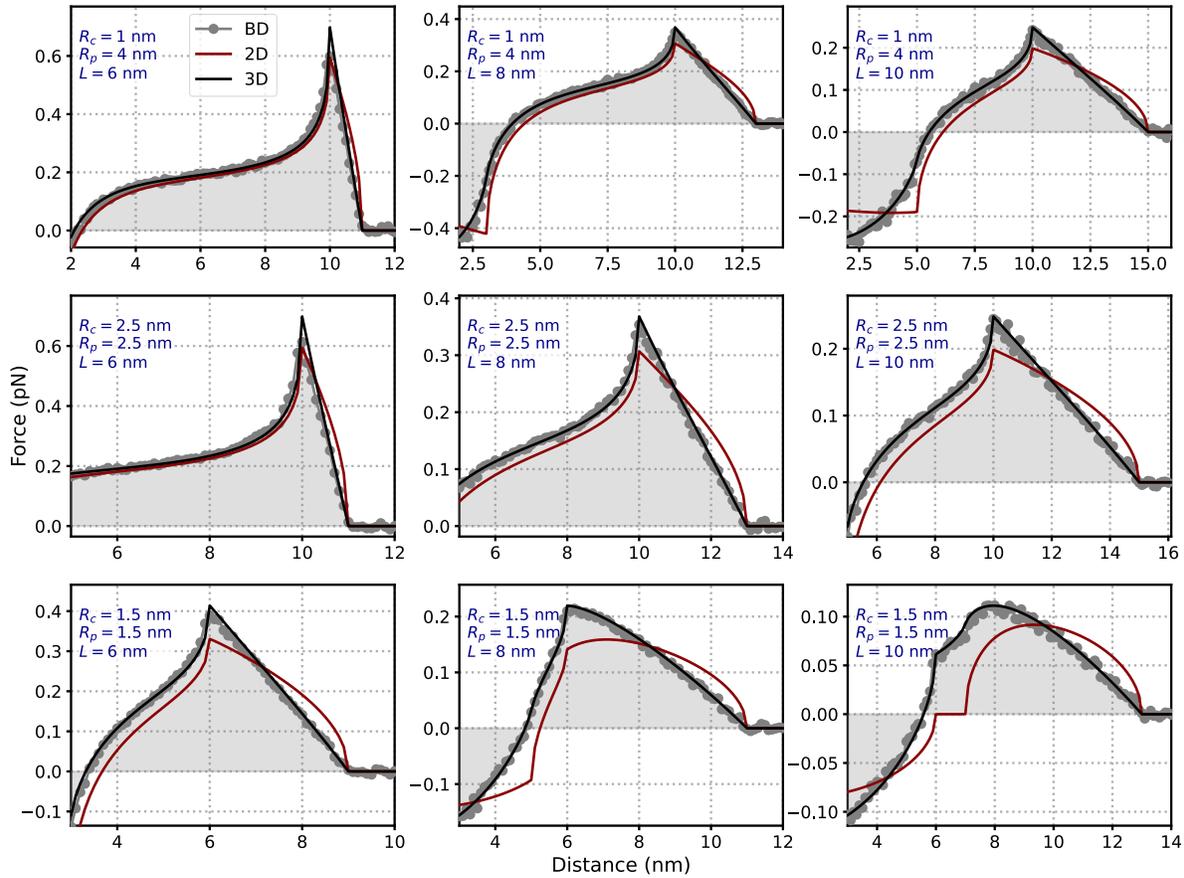

**Figure 5**. Force profiles from Brownian dynamics closely follow the analytical results. Each panel shows the force per particle as a function of the distance for the parameter values indicated in their insets. Brownian dynamics results are shown as grey circles, and analytical 3D calculations are plotted as a black line in each panel. Results for the 2D system are shown as dark red lines for comparison.

The force acting on the filaments is computed as the time average of the net force $\mathbf{F}_i$ exerted by the tethered particles on the filaments. The impulsive character of the hard-core and tethering interactions implies that each collision or constraint event contributes a large force over a short time. This contribution becomes larger as $F_W$ increases, but the duration becomes correspondingly shorter, yielding a consistent value of the average force independent of $F_W$, provided that $F_W$ is sufficiently large.

To characterize the dependence of the entropic force on the inter-filament distance, we computed the forces for a range of distances. Starting from the closest possible distance ($d = 2R_c$), the average force was computed over 100 ns, the separation was increased by 0.1 nm, the system was allowed to equilibrate for an additional 100 ns, and this process was iterated until the entropic force became negligible (typically for $d \gtrsim 2L$). Representative snapshots of the simulated system at different inter-filament distances are shown in Figure 4.

The force profiles obtained for a diverse range of parameter values in the numerical experiments perfectly match those of the exact analytical 3D calculations, including positive and negative values, the non-monotonous behavior, and the continuous but non-differentiable shape (Figure 5). The results for the 2D description are similar to those of the 3D case, especially for $L \simeq R$. In all cases (exact 2D and 3D results and numerical experiments), the system behavior exhibits attractive forces at short distances for sufficiently small values of $R/L$.

## VI. Invariance and scaling laws

Scaling laws provide an avenue to uncover fundamental relations between different quantities and exponents[22]. At the component level, four parameters, $R_c$, $R_p$, $L$, and $d$, describe the system. Entropy and its derived quantities depend only on three parameters: $R = R_c + R_p$, $L$, and $d$. The separation between filaments, however, depends on the explicit value of $R_c$, with the minimal separation being $2R_c$.

As the scaling parameter, we use $L$, which leads to the nondimensional quantities $\tilde{R} = R/L$ and $\tilde{d} = d/L$. The free volume of the particle scales as $V_F(L, R, d) = L^3 V_F(1, \tilde{R}, \tilde{d})$. The corresponding entropy $S(L, R, d) = S(1, \tilde{R}, \tilde{d}) + k_B \ln\left(\frac{L}{\lambda}\right)^3$ does not scale but the entropy difference does with a zero exponent, thus remaining invariant. Consequently, the entropic force scales as $f(L, R, d) = \frac{1}{L} f(1, \tilde{R}, \tilde{d})$. This scaling law dictates that $f(L, R, d)L$ remains invariant when changing the tether length while keeping $\tilde{R}$ and $\tilde{d}$ constant, which leads to forces that decrease with $L$.

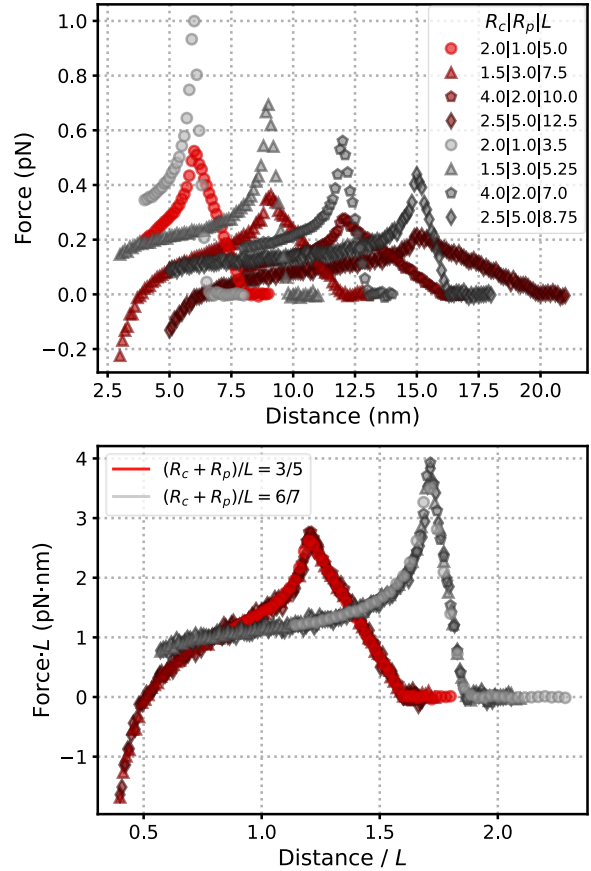

**Figure 6**. Scaling law for force profiles. Multiple force profiles (top) can be scaled into a single curve (bottom) for the force times the tether length as a function of the distance normalized by the tether length. The shape of the curve depends on a single parameter $\tilde{R} = (R_c + R_p)/L$, illustrated for $\tilde{R} = 3/5$ (red symbols) and $\tilde{R} = 6/7$ (grey symbols).

We computed the force profiles for different sets of parameter values, including different values of $L$, $R_c$, and $R_p$ (Figure 6), leading to profiles with sustained non-zero forces that spread over 20 nm with values that range from $-0.25\, pN$ up to $1\, pN$, with an overall decreasing trend as $L$ increases. In contrast, multiple scaled force profiles, $f(1, \tilde{R}, \tilde{d})$ as a function of $\tilde{d}$, collapse into a single curve determined by the value of $\tilde{R}$.

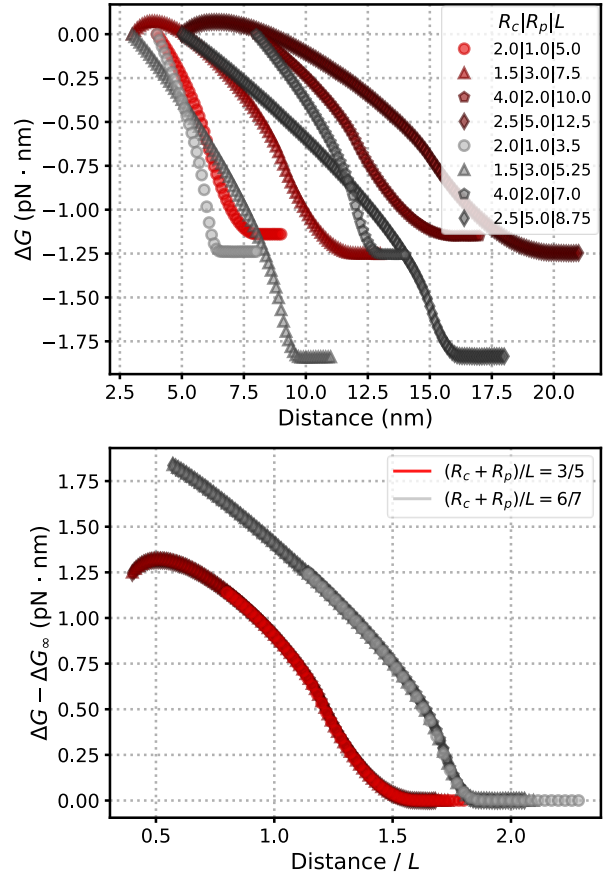

**Figure 7**. Scaling law for free energy profiles. Multiple free energy profiles (top) can be scaled into a single curve (bottom) for the free energy of separation ($\Delta G$) relative to its value at infinite separation ($\Delta G_\infty$) as a function of the distance normalized by the tether length. The shape of the curve depends on a single parameter $\tilde{R} = (R_c + R_p)/L$, illustrated for $\tilde{R} = 3/5$ (red symbols) and $\tilde{R} = 6/7$ (grey symbols).

The free energy of separation does not scale because its reference point at $d = 2R_C$ does not scale. However, it scales relative to its value at infinite separation $\Delta G(L, R, d) - \Delta G(L, R, \infty) = \Delta G(1, \tilde{R}, \tilde{d}) - \Delta G(1, \tilde{R}, \infty)$. Here, we have used that $d = \infty$ is equal to $d = L\infty$ for any tether length different from 0. The free energies do not exhibit an obvious trend (Figure 7). However, multiple free energies relative to their value at infinite separation as functions of $\tilde{d}$ collapse into a single curve determined by the value of $\tilde{R}$.

## VII. Conclusion

Biomolecular processes involved in the disassembly of toxic fibrillar protein aggregates rely on particles tethered to bundled filaments for generating repulsive forces between the filaments. These filaments consist of polymeric stacks of misfolded protein monomers that must be depolymerized to allow for proper refolding or clearance. The purely entropic process described here results in the collective separation of the underlying subsystems, each comprising a filament and its attached particles, just before filament depolymerization[2].

We have shown that the underlying mechanism is much more complex and exhibits richer phenomenology than previously anticipated based on its biological function. Our analytical exact results have revealed large parameter domains in which the net force becomes attractive rather than repulsive. In the face of this inherently intricate behavior, we found general scaling laws that show that the force profiles are determined by one single dimensionless parameter: the excluded-volume-radius-to-tether-length ratio. Ratios close to one lead to higher repulsive entropic forces whereas lower ratios result in larger domains with attractive forces.

The emergence of attractive forces can be understood through two complementary physical pictures. Mechanistically, the behavior is dictated by where the tethered particles can collide. When tethers are short ($R/L \approx 1$), particles are confined near the filaments, resulting in near-side collisions that push the filaments apart (repulsion). However, when tethers are sufficiently long ($R/L$ small), particles can reach around and collide with the far side of the opposing filament. These far-side impacts effectively pull the filaments together, generating a net attractive force. Thermodynamically, this corresponds to an entropic effect analogous to depletion forces. When the filaments approach each other, their excluded volumes (the regions inaccessible to the particles) begin to overlap. This overlap increases the total free volume available to the tethered particles, thereby increasing the system's entropy. Since the system seeks to maximize entropy, this generates a thermodynamic force driving the filaments into the overlapped (attractive) state.

The detailed characterization that we have provided here is needed not only for devising potential therapeutic strategies for the underlying neuropathologies but also for guiding the design of new artificial nanosystems with specific properties. In this regard, the behavior can be tuned by selecting particle sizes and tether lengths, both of which can be constructed through chemical synthesis or molecular biology tools[17,23]. Therefore, our results provide essential information for a wide range of applications, including fields as diverse as biophysics[15], soft matter physics[16], targeted drug delivery[17], and nanoparticle assembly[18].

## Acknowledgments

J.M.G.V. acknowledges support from Ministerio de Ciencia, Innovación y Universidades (MICIU/AEI/10.13039/501100011033/FEDER, UE) under Grant No. PID2024-160016NB-I00.